\documentclass{article}
\usepackage{spconf,amsmath,graphicx,hyperref}
\usepackage{cite}
\usepackage{amssymb,amsfonts}
\usepackage{algorithmic}
\usepackage{textcomp}
\usepackage{xcolor}
\def\BibTeX{{\rm B\kern-.05em{\sc i\kern-.025em b}\kern-.08em
    T\kern-.1667em\lower.7ex\hbox{E}\kern-.125emX}}
\usepackage{xcolor}
\usepackage{pifont}
\newcommand{\cmark}{\ding{51}}
\newcommand{\xmark}{\ding{55}}
\usepackage{times}
\usepackage{soul}
\usepackage{url}
\usepackage[utf8]{inputenc}
\usepackage{caption}
\usepackage{graphicx}
\usepackage{amsmath}
\usepackage{amsthm}
\usepackage{booktabs}
\usepackage{algorithm}
\usepackage{algorithmic}
\usepackage[switch]{lineno}
\usepackage{amsfonts}
\usepackage{subcaption}

\title{Graph-Based Learning of Spectro-Topographical EEG Representations with Gradient Alignment for Brain-Computer Interfaces}
\name{Prithila Angkan$^{1,2}$, Amin Jalali$^{1,2}$, Paul Hungler$^{1,3}$, Ali Etemad$^{1,2}$}
\address{$^{1}$Ingenuity Labs Research Institute \\
$^{2}$ Department of Electrical and Computer Engineering \\ $^{3}$Department of Chemical Engineering \\
Queen's University, Kingston, Canada\\
\{prithila.angkan, amin.jalali, paul.hungler, ali.etemad\}@queensu.ca}

\begin{document}
%
\maketitle
\begin{abstract}
We present a novel graph-based learning of EEG representations with gradient alignment (GEEGA) that leverages multi-domain information to learn EEG representations for brain-computer interfaces. Our model leverages graph convolutional networks to fuse embeddings from frequency-based topographical maps and time-frequency spectrograms, capturing inter-domain relationships. GEEGA addresses the challenge of achieving high inter-class separability, which arises from the temporally dynamic and subject-sensitive nature of EEG signals by incorporating the center loss and pairwise difference loss. Additionally, GEEGA incorporates a gradient alignment strategy to resolve conflicts between gradients from different domains and the fused embeddings, ensuring that discrepancies, where gradients point in conflicting directions, are aligned toward a unified optimization direction. We validate the efficacy of our method through extensive experiments on three publicly available EEG datasets: BCI-2a, CL-Drive and CLARE. Comprehensive ablation studies further highlight the impact of various components of our model.
\end{abstract}
\section{Introduction}
Electroencephalography (EEG) is a non-invasive technique that captures the electrical activity of the brain. Its cost-effectiveness and high temporal resolution make it widely used for brain-computer interfaces (BCI) in various research areas \cite{zheng2024semi,ding2024eeg,grover2024segment}. However, EEG presents challenges due to its low signal-to-noise ratio, subject-dependency, and low spatial resolution \cite{he2019transfer}.  
Prior EEG studies leverage information from various domains such as time, frequency, and topographical mapping to enhance representations \cite{li2021multi,yao2024emotion}. However, learning effective multi-domain representations from EEG poses two nuanced challenges. First, obtaining distinct class-specific clusters with large inter-class separation has proven challenging, especially in multi-domain setups \cite{yamamoto2023modeling}. Second, to learn multi-domain information, gradient conflicts can arise, resulting in suboptimal training \cite{wei2024mmpareto}.

To address these challenges, we propose a novel approach using \textbf{G}raph-based learning of spectro-topographical \textbf{EE}G representations with \textbf{G}radient \textbf{A}lignment (GEEGA). GEEGA encodes EEG from frequency-based topography maps and time-frequency spectrograms, maps embeddings onto a shared feature space using graph convolutional networks, and aligns gradients to reduce domain conflicts. 
Our method calculates class centers and pulls positive pairs toward them while pushing negatives apart for maximum inter-class separation. We evaluate our method on three publicly available EEG datasets, CLARE \cite{bhatti2024clare}, CL-Drive \cite{angkan2024multimodal}, and BCI-2a \cite{brunner2008bci}. Our approach achieves state-of-the-art performance across all three benchmarks.

The contributions in this work are summarized as follows.
    (\textbf{1}) We propose a new model, GEEGA, for EEG representation learning. Our model successfully learns multi-domain spectro-topographical information from EEG through graph-based fusion. 
    (\textbf{2}) Our model effectively resolves gradient conflicts by aligning the gradients of the fused embeddings, ensuring that discrepancies, where gradients from each domain point in different directions, are addressed and guided toward a unified direction. This ensures balanced optimization across all domains causing the fused embeddings effectively capture complementary information from different domains, leading to enhanced performance. To the best of our knowledge, this is the first attempt to resolve gradient conflicts in the context of BCI as well as the first effort toward addressing such conflicts in a \textit{multi-domain} setting in any context.
    (\textbf{3}) Moreover, our model incorporates class centers, enhancing inter-class separability by pulling positive pairs toward their respective class centers while pushing negative pairs apart.
    (\textbf{4}) GEEGA shows strong performances across several datasets and outperforms prior works. Detailed ablation studies demonstrate the positive impact of different components of our method.

\section{Related Work}
Transformers have recently become popular in EEG representation learning. In \cite{ding2024eeg}, EEG-Deformer was proposed combining CNNs with transformers to capture coarse and fine-grained temporal dynamics. In \cite{yao2024emotion} parallel transformers were used for spatial-temporal feature extraction with CNN integration, while \cite{wan2023eegformer} employed CNNs for channel-wise feature extraction followed by transformer processing.
EEG channel-attention with Swin Transformer for motor imagery was integrated in \cite{wang2023novel} and \cite{xu2023amdet} and utilized multi-dimensional global attention for spectral-spatial-temporal features. In \cite{pulver2023eeg} self-supervised masked autoencoders for cognitive load classification were applied, while \cite{liu2023bstt} implemented Bayesian transformers for sleep staging.

Graph-based architectures have gained traction for EEG classification. GCN was used in \cite{jia2020graphsleepnet} for sleep stage classification to learn intrinsic channel connections. In \cite{lin2023eeg}, graph and 1D convolutions were combined for intra- and inter-channel interactions, while \cite{gu2023domain} integrated GCNs with LSTMs for emotion classification. GCN and attention mechanisms were fused in \cite{jin2024pgcn} for structural relationships and long-range dependencies. Another graph-based network was used in \cite{liu2024vsgt}, leveraging the spatial and temporal dependencies of EEG for emotion recognition. Finally \cite{song2020instance} dynamically adjusted graph connections per instance using multi-level graph convolutions and coarsening.

\begin{figure*}[htp]
\centering
\begin{tabular}{@{}c@{}}
\subfloat{\includegraphics[width=0.63\linewidth]{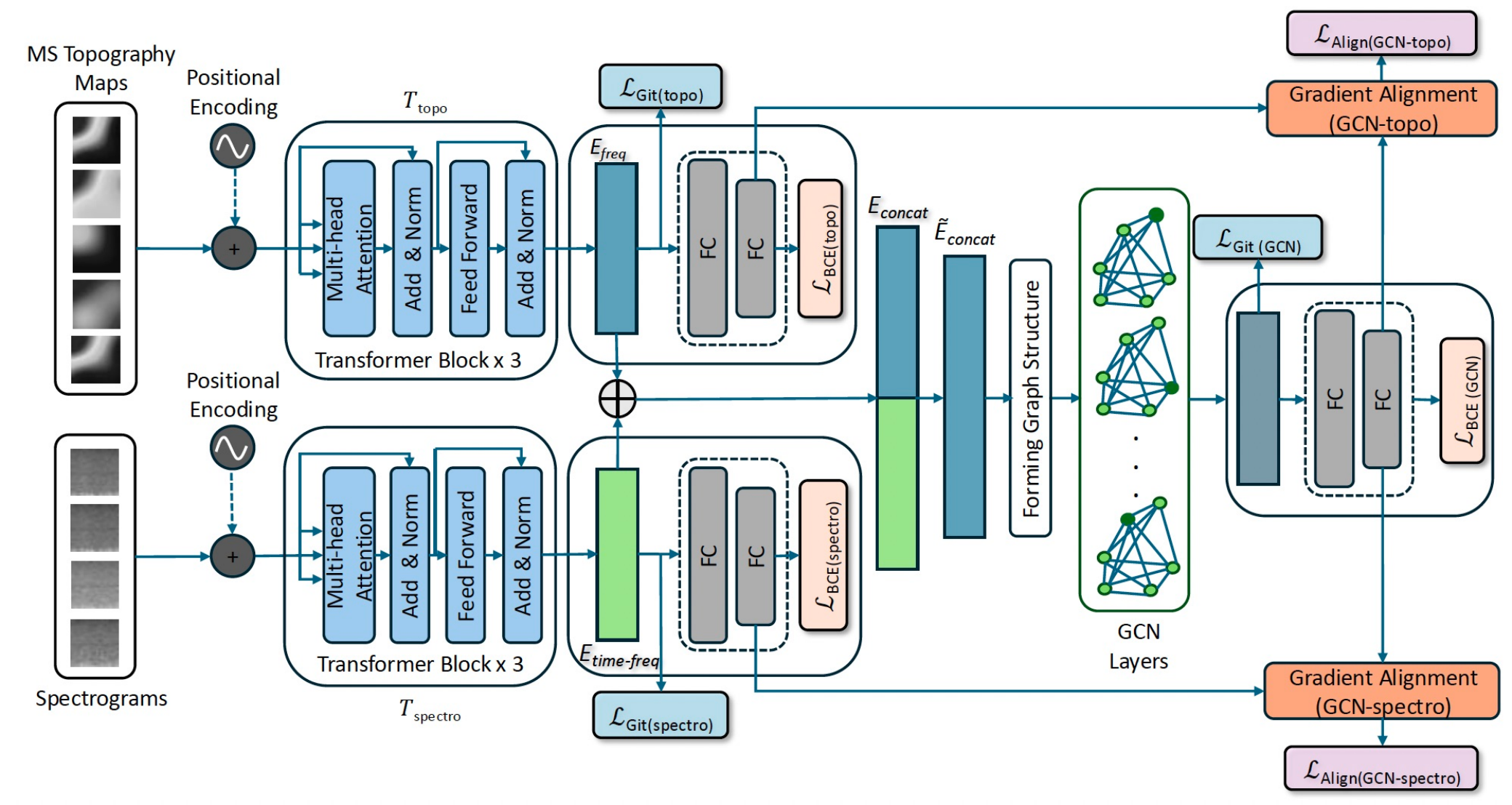}} \\ 
\small (a) Proposed model.
\end{tabular}
\qquad
\begin{tabular}{@{}c@{}}
\subfloat{\includegraphics[width=0.23\linewidth]{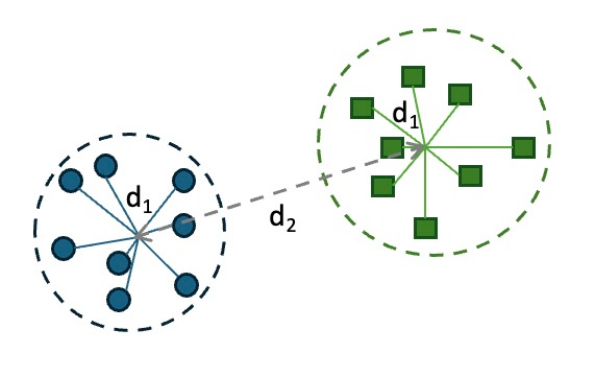}} \\ 
\small (b) Git loss. \\[0.1cm]
\subfloat{\includegraphics[width=0.32\linewidth]{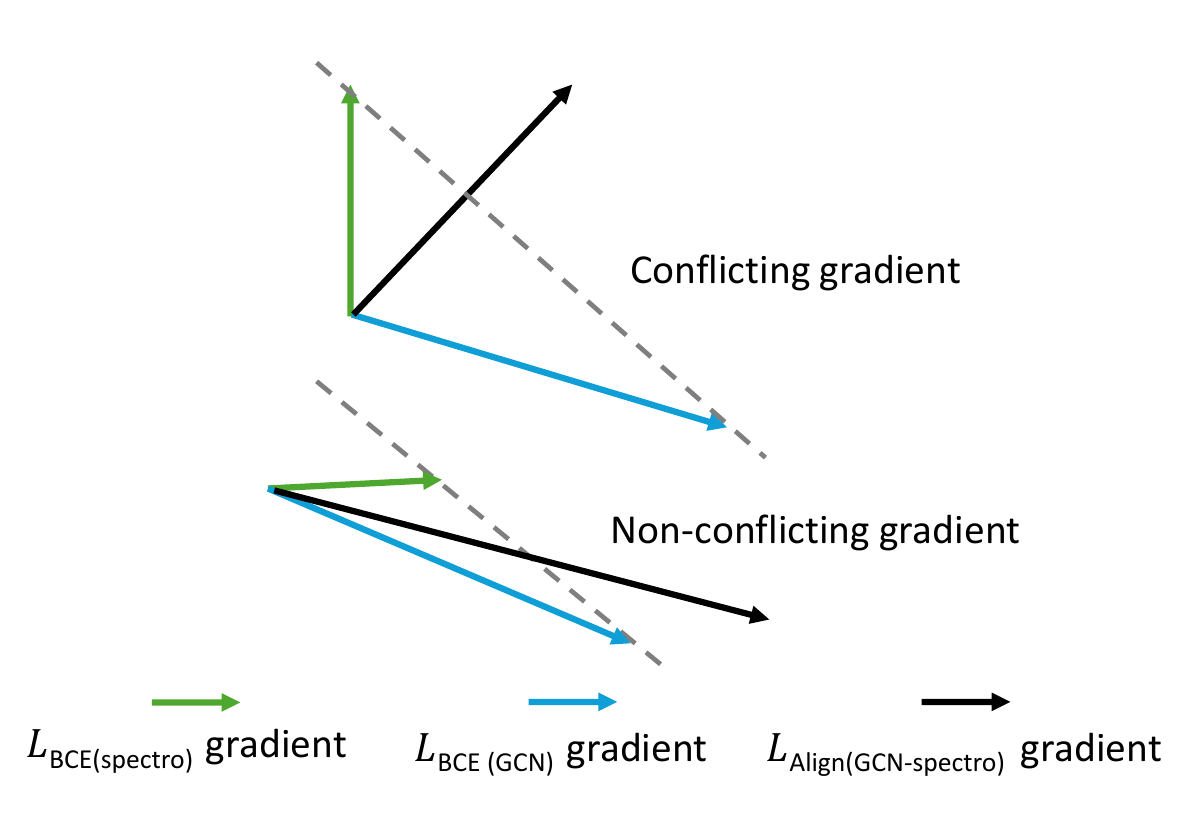}} \\ 
\small (c) Gradient alignment. 
\end{tabular}
\vspace{-3mm}
\caption{(a) The overview of our proposed network is depicted. (b) The concept of the Git loss is presented where we aim to minimize intra-class distances $d_{1}$ and maximize inter-class distances $d_{2}$. (c) The concept of gradient alignment is presented.}
\label{fig:model}
\end{figure*}

\section{Method}\label{Method}

\subsection{Problem Statement}
Given a set of EEG signals, $X = [X_1, X_2, \cdots, X_c] \in \mathbb{R}^{c}$ with $c$ channels, we aim to extract complementary representations: frequency domain $E_{\text{freq}} \in \mathbb{R}^{\text M_1}$ and time-frequency domain $E_{\text{time-freq}} \in \mathbb{R}^{\text M_2}$, where $M_1$ and $M_2$ are the size of the embeddings. Training a unified multi-domain model faces the challenge of misaligned gradients. Specifically, the gradients measured by the loss function over a mini-batch $B$ for the frequency domain ($\nabla_\text B^{\text{freq}}$), for the time-frequency domain ($\nabla_\text{B}^{\text{time-freq}}$), and the fused domain ($\nabla_ \text{B}^{\text{joint}}$), often point to conflicting directions, hindering effective training. Our goal is to align these gradients for unified optimization while achieving high inter-class separability.

\subsection{Our Approach}
\noindent\textbf{Multi-domain encoding.}
We encode the pre-processed EEG signals $X$ into multi-spectral topography maps $X_{\text{topo}} \in \mathbb{R}^{B \times k \times h \times w}$ (frequency domain) and spectrograms $X_{\text{spectro}} \in \mathbb{R}^{B \times c \times h \times w}$ (time-frequency domain), where $B$, $k$, $c$, $h$, $w$ denote batch size, frequency bands, channels, height, and width respectively. Both inputs are flattened, linearly projected into token sequences \cite{dosovitskiy2021an}, and positional encoding is added. The tokens are then fed to their respective transformer branches: $T_{\text{topo}}$ (frequency domain encoding) and $T_{\text{spectro}}$ (time-frequency domain encoding). Producing embeddings $E_{\text{freq}} \in \mathbb{R}^{M_1}$ (frequency domain) and $E_{\text{time-freq}} \in \mathbb{R}^{M_2}$ (time-frequency domain), where $M_1$ and $M_2$ denote the size of the embeddings (see Fig. \ref{fig:model} (a)).


\noindent\textbf{Graph-based embedding fusion.}
We fuse the embeddings $E_{\text{freq}}$ and $E_{\text{time-freq}}$ using a GCN module $\Phi$. The concatenated embedding $E_{\text{concat}} \in \mathbb{R}^{B \times G_1}$ is projected to $\tilde{E}_{\text{concat}} \in \mathbb{R}^{B \times G_2}$ where $B$ is the batch size, $G_1$ is the initial embedding dimension, and $G_2$ is dimension of the higher-dimensional space, which is defined as $G_2=N \times F$, where $N$ is the number of nodes in the graph with $F$ being the feature dimension of each node. $\tilde{E}_{\text{concat}}$ is reshaped into $\tilde{E}_{\text{node}} \in \mathbb{R}^{B \times N \times F}$ to form a graph structure.

In the first GCN layer, the learnable weight matrix ${W}_1 \in \mathbb{R}^{F \times F}$ transforms the node features as: 
\begin{equation}
O_{\text{GCN}_\text{1}}= \tilde{{E}}_{\text{node}} {W}_1, \quad O_{\text{GCN}_\text{1}} 
\in \mathbb{R}^{B \times N \times F},
\end{equation}
where $O_{\text{GCN}_\text{1}}$ is the output from the first GCN layer.
Node features are updated by aggregating neighboring information via adjacency matrix $A \in \mathbb{R}^{N \times N}$, forming a fully connected graph in our case as:
\begin{equation}
\tilde{{E}}_{\text{node-update}} = {A} \cdot O_{\text{GCN}_\text{1}} .    
\end{equation}
This process is repeated for the second GCN layer, followed by flattening and a linear transformation to produce the final feature vector of size $H$. A ReLU activation function is applied after each GCN layer to introduce non-linearity.

To train $T_{\text{topo}}$, $T_{\text{spectro}}$, and the GCN, we use binary cross-entropy loss $\mathcal{L}_{\text{BCE}}$ and Git loss \cite{calefati2018git}. 
Git loss is defined as:
\begin{equation}
\mathcal{L}_{\text{Git}} = \frac{1}{2} \sum_{i=1}^n \|E^{i} - c_{y}^{i}\|_2^2 + \sum_{i,j=1, i \neq j}^m \frac{1}{1 + \|E^{i} - c_{y}^{j}\|_2^2},
\end{equation}
where $E^i$ is the feature vector of the $i^{th}$ sample, and $c_{y}^{i}$ is the center of the class to which $E^i$ belongs.  $n$ and $m$ are the total number of samples for the two classes, respectively. 
This loss combines center loss (first term of the equation) which reduces intra-class distances with pairwise difference loss (second part of the equation) which increases inter-class distances to enhance class separability as shown in Fig.\ref{fig:model} (b).

\noindent \textbf{Gradient alignment.}
Multiple domains in a single latent space can face the \textit{gradient conflict} problem where the gradients from different domains may point at conflicting directions \cite{peng2022balanced,wei2024mmpareto} (see Fig.\ref{fig:model}(c)). This can result in sub-optimal training of the model and degrading of downstream performance. We align the two domains with respect to the fused domain rather than directly aligning the individual domains with each other as non-linear fusion reveals complex cross-domain interactions that remain hidden when domains are considered in isolation \cite{ghosh2024nonlinear}.

We define the gradients of losses computed over a mini-batch $B$ as $\nabla_B \mathcal{L}_{\text{BCE(topo)}}$, $\nabla_B \mathcal{L}_{\text{BCE(spectro)}}$, and $\nabla_B \mathcal{L}_{\text{BCE(GCN)}}$. When cosine similarity between gradients is negative ($\cos \beta \leq 0$, where $\beta$ represents the angle between the gradients from different domains), conflicts exist. To resolve this, we use the Pareto optimization method that assigns weights $\alpha^{\text{topo}}$, $\alpha^{\text{spectro}}$, and $\alpha^{\text{GCN}}$ via a closed-form solution. The optimization problem for aligned gradient $\mathcal{L}_{\text{Align(GCN-topo)}}$ is 
\begin{equation} \label{eq:pareto}
\min_{\alpha^{\text{GCN}}, \alpha^{\text{topo}} \in \mathbb{R}} \left\| \alpha^{\text{GCN}} \nabla_B \mathcal{L}_{\text{GCN}} + \alpha^{\text{topo}} \nabla_B \mathcal{L}_{\text{topo}} \right\|^2,
\end{equation}
subject to the constraints that
$\alpha^{\text{GCN}}, \alpha^{\text{topo}} \geq 0$ 
and
$\alpha^{\text{GCN}} + \alpha^{\text{topo}} = 1$.
Here, Eq. \ref{eq:pareto} minimizes the $L_2$-norm of the gradients within the convex hull of the gradient vectors $\{ \nabla_B \mathcal{L}_i \}_{i \in \{\text{GCN}, \text{topo}\}}$ \cite{desideri2012multiple}.
The aligned gradient is:
\begin{equation}
h_{\text{GCN-topo}}^{\text{align}}(\theta) = 2\alpha^{\text{GCN}} \nabla_B \mathcal{L}_{\text{GCN}}(\theta) + 2\alpha^{\text{topo}} \nabla_B \mathcal{L}_{\text{topo}}(\theta),    
\end{equation}
where the resulting weights $2\alpha^{\text{GCN}}$ and  $2\alpha^{\text{topo}}$ maintain the same weight summation (i.e., $2\alpha^{\text{GCN}} + 2\alpha^{\text{topo}} = 2$) and the model parameters $\theta$ are updated as
\begin{equation}
\theta(t + 1) = \theta(t) - \eta h_{\text{GCN-topo}}^{\text{align}}(\theta(t)).    
\end{equation}
Similar operations are performed for $\mathcal{L}_{\text{Align(GCN-spectro)}}$ to align GCN and spectrogram gradients. 

Finally, we define the total loss of GEEGA as:
\begin{equation}
\begin{aligned}
       \mathcal{L}_{Total} = & \mathcal{L}_{\text{Git(topo)}} + \mathcal{L}_{\text{Git(spectro)}} + \mathcal{L}_{\text{Git(GCN)}}\\ 
   & +\mathcal{L}_{\text{BCE(topo)}} + \mathcal{L}_{\text{BCE(spectro)}} + \mathcal{L}_{\text{BCE(GCN)}}\\
   & +\mathcal{L}_{\text{Align(GCN-topo)}}  + \mathcal{L}_{\text{Align(GCN-spectro)}} .
\end{aligned}
\end{equation}

\section{Experiment setup}\label{Experimentation}
\noindent \textbf{Datasets.} \label{Datasets}
We use three publicly available EEG datasets, namely BCI-2a \cite{brunner2008bci}, CL-Drive \cite{angkan2024multimodal} and CLARE \cite{bhatti2024clare} for our work. We use leave-one-subject-out (LOSO) evaluation. 
For BCI-2a, feet and tongue movement are used for binary classification, while for CL-Drive and CLARE, the subjective scores are binarized into low (1-5) and high (6-9) categories. 





\noindent \textbf{Data preprocessing.}
For BCI-2a, we use pre-processed data with each trial as an individual segment. For the other two datasets, we apply Butterworth bandpass filtering (1-75 Hz) and notch filtering following \cite{angkan2024multimodal}, then segment the signals into 10-second intervals. We generate multi-spectral topography maps and spectrograms from the segmented data. 


\noindent \textbf{Multi-spectral topography maps.}
To generate multi-spectral topography maps, we compute power spectral density (PSD) for each channel and five frequency bands: Delta, Theta, Alpha, Beta, and Gamma, following standard EEG practice \cite{song2018eeg,wang2024dmmr,angkan2024multimodal}. Using Simpson's rule \cite{velleman2005generalized}, we compute each band's power across all channels. These values are spatially mapped onto 2D grids using the international 10-20 electrode system with radial basis function (RBF) interpolation \cite{havugimana2023deep}, creating multi-spectral topography maps of dimensions $32 \times 32 \times 1$ for all datasets.


\noindent \textbf{Spectrograms.}
While PSD captures power distribution across frequency bands, it fails to capture temporal dependencies. We address this using spectrograms containing time-frequency information. We compute Fast Fourier Transform (FFT) with non-overlapping 256-point windows, creating matrices where columns represent frequencies and rows represent time intervals. 
Spectrograms are generated for 4 channels (cognitive load datasets) or 22 channels (motor imagery dataset), each with dimensions $32 \times 32 \times 1$


\begin{table}[t]
    \caption{Performance compared to state-of-the-art solutions.}
    \label{table:comparison}
    \small
    \setlength
    \tabcolsep{2pt}
        \begin{center}{
\resizebox{1\linewidth}{!}{  
            \begin{tabular}{l|ll|ll|ll}
                 \hline
            \multicolumn{1}{c|}{} &
            \multicolumn{2}{c|}{BCI-2a}& \multicolumn{2}{c|}{CL-Drive}&
            \multicolumn{2}{c}{CLARE}\\
                 Model   & Accuracy & F1 &  Accuracy & F1 &
                 Accuracy & F1\\
                 \hline\hline
                 DGCNN \cite{song2018eeg} & 65.29(9.26) & 64.74(11.82) & 65.77(4.71) & 57.06(5.30) & 61.84(3.96) & 51.05(7.70)\\
                 BiHDM \cite{li2020novel} & 67.86(9.29) & 67.27(10.57) & 62.01(15.57) & 57.92(11.66) & 68.14(16.43) & 52.17(16.54)\\
                 Conformer \cite{song2022eeg} & 68.12(9.43) & 67.53(11.25) & 69.38(8.72) & \underline{63.29(9.29)} & \underline{70.42(16.02)} & 58.28(12.00)\\ 
                 MAE \cite{pulver2023eeg}  & 65.76(10.24) & 65.98(10.92) & 67.88(14.67) & 61.25(13.18) & 62.48(10.71) & 57.51(7.29)\\
                 VGG-style \cite{angkan2024multimodal} & \underline{69.48(10.67)} & \underline{69.73(10.24)} & \underline{70.28(10.87)} & 63.12(9.39) & 70.29(16.03) & \underline{60.24(13.16)}\\
                 DMMR \cite{wang2024dmmr} & 65.57(10.23) & 64.97(10.20) & 61.15(13.74) & 52.40(8.28) & 69.02(22.07) & 52.95(14.71)\\
                 \textbf{GEEGA (our)}    & \textbf{73.54(8.66)} & \textbf{72.86(8.04)} & \textbf{74.64(7.56)} & \textbf{64.53(8.24)} & \textbf{73.29(16.23)} & \textbf{60.68(14.42)}\\
                 \hline
                \end{tabular} 
                }
                }
        \end{center}
\end{table}

\begin{table}[t]
    \caption{Ablation experiments demonstrating the impact of each module within our proposed model. MS: multi-spectral topography maps, S: spectrograms, A: alignment.}
    \label{table:ablation}
    \small
    \setlength
    \tabcolsep{2pt}
        \begin{center}{
            \resizebox{1\linewidth}{!}{\begin{tabular}{cccc|ll|ll|ll}
                 \hline
             & & & & 
            \multicolumn{2}{c|}{BCI-2a}& \multicolumn{2}{c|}{CL-Drive} &
            \multicolumn{2}{c}{CLARE}\\
                 MS    & S &$\mathcal{L}_{Git}$  & A & Accuracy & F1 & Accuracy & F1 &
                 Accuracy & F1\\
                 \hline\hline
                 \cmark &\cmark  &\cmark  &\cmark & \textbf{73.54(8.66)} & \textbf{72.86(8.04)} & \textbf{74.64(7.56)} & \textbf{64.53(8.24)} & \textbf{73.29(16.23)} & \textbf{60.68(14.42)}\\ 
                 \cmark &\cmark &\xmark &\cmark & 70.85(9.24) & 69.20(9.83) & 69.30(10.38) & 60.07(7.72) & 69.41(15.84) & 54.28(12.30) \\
                 \cmark &\cmark &\cmark &\xmark & 70.90(9.45) & 69.87(9.73) & 72.70(8.47) & 62.65(6.81) & 71.05(16.50) & 56.39(14.00)\\              
                 \cmark &\cmark  &\xmark &\xmark & 69.48(8.29) & 68.21(8.84) & 70.20(8.99) & 60.29(6.43) & 70.07(16.50) & 56.40(14.00)\\
                 \cmark &\xmark  &\xmark  &\xmark & 66.00(9.41) & 65.15(9.38) & 67.52(9.29) & 60.23(7.25) & 66.82(17.24) & 54.52(14.69)\\
                 \xmark &\cmark  &\xmark  &\xmark & 66.43(9.36) & 64.92(8.25) & 67.46(8.57) & 59.43(7.34) & 70.50(15.38) & 52.47(17.16)\\
                 \hline
                \end{tabular} 
                }
                }
        \end{center}
\end{table}

\noindent \textbf{Implementation details.}
We use a batch size $B$ of 32 and the Adam optimizer \cite{adam} (learning rate 0.0001, weight decay 0.00001). Training employs a Plateau scheduler (decay factor 0.1, patience 5) and warmup LambdaLR for first 5 epochs. Model is trained for 25 epochs on NVIDIA 2080 Ti using PyTorch. Both encoders ${T}_{\text{topo}}$ and ${T}_{\text{spectro}}$ use 3 transformer blocks with 8 attention heads, embedding dimension 512, and MLP hidden dimension 1024. The GCN module parameters are: $G_1$ = 1024, $G_2$ = 1536, $N$ = 6 nodes, $F$ = 256 and $H$ = 512. This connects to FC layers (128, 1) with ReLU activation and 0.25 dropout. The FC layers after $E_{\text{freq}}$ and $E_{\text{time-freq}}$ use identical configurations. Dropout rates of 0.1 and 0.25 are applied to transformer/GCN blocks and FC layers respectively for regularization.

\noindent \textbf{Baseline methods.}
We compare our proposed method with other popular and state-of-the-art recent works in EEG-based classification, and exclude methods requiring large-scale pre-training (EEGPT \cite{wang2024eegpt}, BENDR \cite{kostas2021bendr}) following \cite{liu2024vsgt,ni2024dbpnet}.

\section{Results}\label{results}  
\noindent \textbf{Performance.} 
We present the overall performance of our method in comparison to prior works in Table \ref{table:comparison}, where we observe that GEEGA achieves the best result across all three datasets. 
Notably, we observe that our method achieves higher accuracy and F1 scores than the two competing methods, the VGG-style CNN \cite{angkan2024multimodal} and Conformer \cite{song2022eeg}, by considerable margins. For instance, GEEGA outperforms the VGG by accuracy and F1 values of 4.06\% and 3.13\% respectively on the BCI-2a dataset, 4.36\% and 1.41\% on the CL-Drive dataset, and 3.00\% and 0.44\% on the CLARE dataset. Similarly, our method outperforms the widely used Conformer model by accuracy and F1 values of 5.42\% and 5.33\% on BCI-2a dataset, 5.26\% and 1.24\% on CL-Drive, and 2.87\% and 2.40\% on CLARE. The results show that performance does not always correlate with the number of parameters, for instance, simpler models like VGG can still perform well.

\begin{figure}[t]
    \begin{subfigure}[t]{0.48\linewidth}
        \centering
        \includegraphics[width = 1\linewidth]{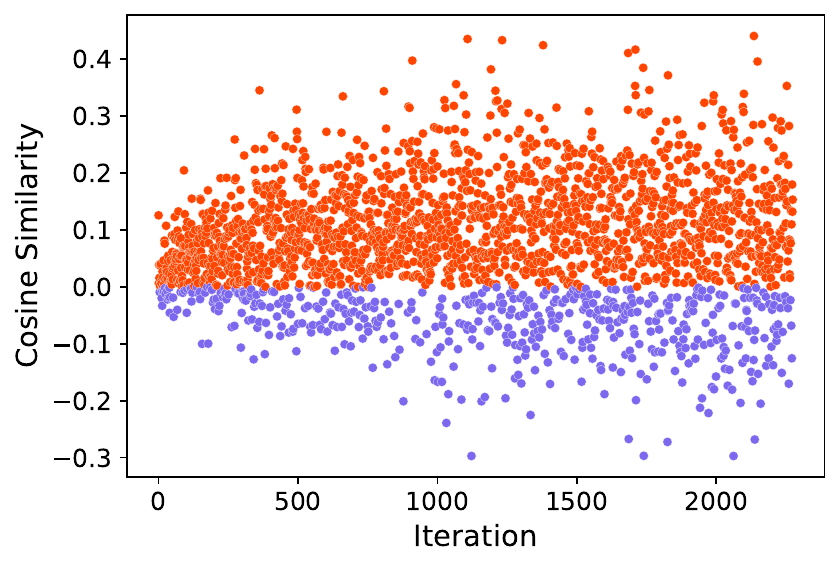}
    \end{subfigure}
    \begin{subfigure}[t]{0.48\linewidth}
        \centering
        \includegraphics[width = 1\linewidth]{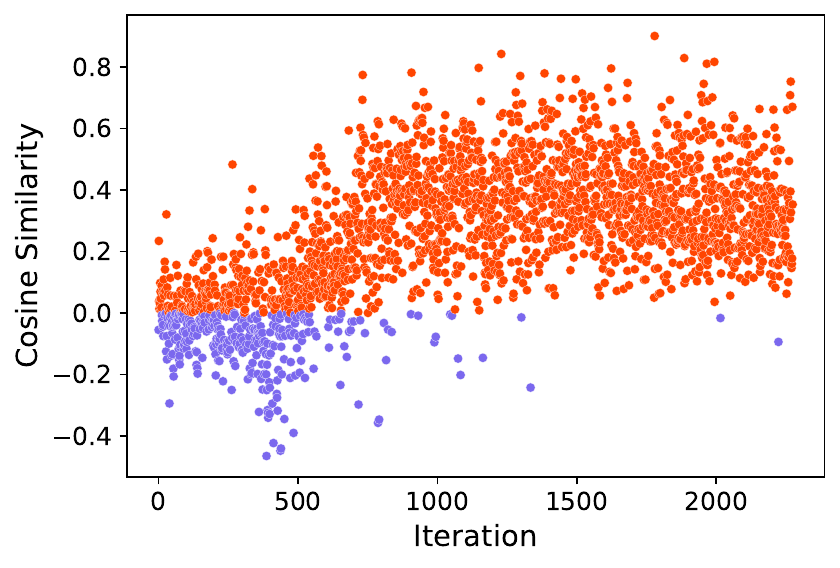}
    \end{subfigure}    
    \centering
    \begin{subfigure}[t]{0.48\linewidth}
        \centering
        \includegraphics[width = 1\linewidth]{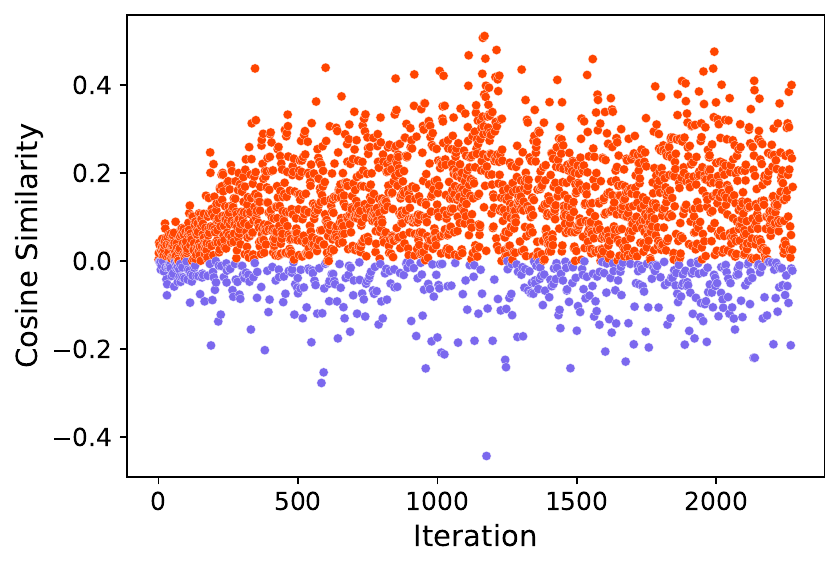}
        \caption{w/o gradient alignment}
    \end{subfigure}
    \centering
    \begin{subfigure}[t]{0.48\linewidth}
        \centering
        \includegraphics[width = 1\linewidth]{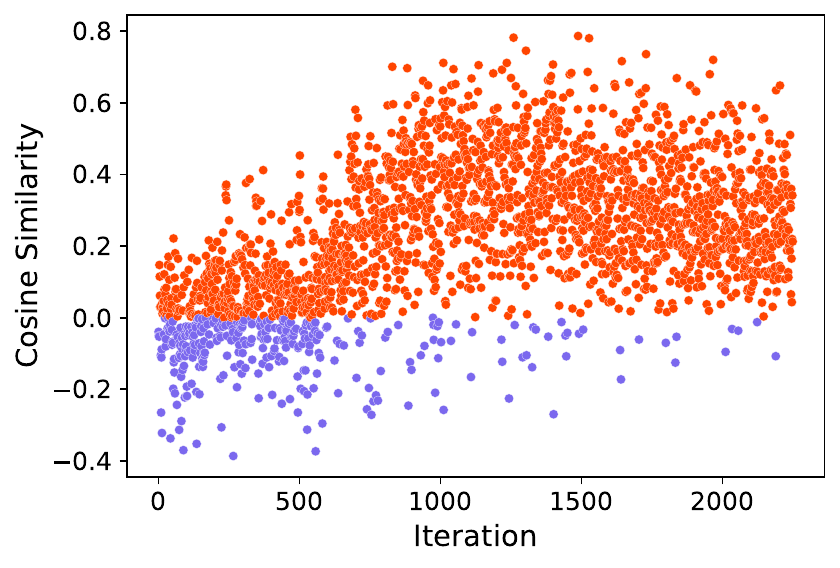}
        \caption{w/ gradient alignment}
    \end{subfigure}
    \centering
        \centering
    \caption{
    The first row shows cosine similarities between multi-spectral topography maps and the fused domain, while the second row shows the same for spectrograms. Blue (values \textless 0) indicates gradient conflicts, while red (values \textgreater 0) indicates no conflict.}
    
\label{fig:cosine_similarity_comparison}
\end{figure}

\noindent \textbf{Gradient alignment.} 
Our key contribution is gradient alignment across domains to minimize conflicts and improve training. Fig. \ref{fig:cosine_similarity_comparison}(a) shows misaligned gradients (positive (red) and negative (blue)) throughout training w/o our alignment process. Fig. \ref{fig:cosine_similarity_comparison} (b) demonstrates reduced misaligned gradients as training progresses, confirming our alignment strategy's effectiveness for both frequency and time-frequency domains.

\noindent \textbf{Ablation.} 
In Table \ref{table:ablation}, we present the results of detailed ablation experiments conducted to evaluate the impact of individual components in our method. We remove key components, including multi-spectral topography maps, spectrograms, the git loss ($\mathcal{L}_{\text{Git}}$), and the alignment mechanism, and compare the results.
We observe that our proposed GEEGA method with all the components achieves the best results compared to the other ablated combinations. Specifically, we observe that removing the git loss or the alignment step individually results in considerable drops in performance.


\section{Conclusion}
\textcolor{black}{We propose GEEGA for EEG representation learning by integrating frequency and time-frequency domains using parallel transformer encoders and graph-based fusion. Our method addresses gradient conflicts through alignment strategies and enhances class separability using center loss with pairwise difference loss. Results on three benchmark datasets demonstrate superior performance over existing methods. In the future cross-task transferability and real-time applications can be explored.
}



\footnotesize
\bibliographystyle{IEEEbib}
\bibliography{refs}

\end{document}